\documentclass[pdflatex,sn-nature, referee]{sn-jnl}

\usepackage{graphicx}
\usepackage{subcaption}
\usepackage{overpic}

\usepackage{graphicx}%
\usepackage{multirow}%
\usepackage{amsmath,amssymb,amsfonts}%
\usepackage{amsthm}%
\usepackage{mathrsfs}%
\usepackage[title]{appendix}%
\usepackage{xcolor}%
\usepackage{textcomp}%
\usepackage{manyfoot}%
\usepackage{booktabs}%
\usepackage{algorithm}%
\usepackage{algorithmicx}%
\usepackage{algpseudocode}%
\usepackage{listings}%

\usepackage{lineno} 

\theoremstyle{thmstyleone}%
\theoremstyle{thmstyletwo}%

\theoremstyle{thmstylethree}%

\begin{document}
\title{Optimizing stellarators with hidden symmetry}


\author[1]{\fnm{Hengqian} \sur{Liu}}
\equalcont{These authors contributed equally to this work.}

\author[1]{\fnm{Guodong} \sur{Yu}}
\equalcont{These authors contributed equally to this work.}

\author*[1]{\fnm{Caoxiang} \sur{Zhu}}\email{caoxiangzhu@ustc.edu.cn}

\author[2]{\fnm{Jos\'e Luis} \sur{Velasco}}

\author[3]{\fnm{Rahul} \sur{Gaur}}

\author[3]{\fnm{Dario} \sur{Panici}}

\author[3]{\fnm{Egemen} \sur{Kolemen}}

\author[4]{\fnm{Mingyang} \sur{Yu}}

\author[1]{\fnm{Weixing} \sur{Ding}}

\author[1]{\fnm{Shaojie} \sur{Wang}}

\author[1]{\fnm{Ge} \sur{Zhuang}}

\affil*[1]{\orgdiv{CAS Key Laboratory of Frontier Physics in Controlled Nuclear Fusion, School of Nuclear Science and Technology}, \orgname{University of Science and Technology of China}, \orgaddress{\city{Hefei}, \postcode{230026}, \state{Anhui}, \country{China}}}

\affil[2]{\orgdiv{Laboratorio Nacional de Fusión}, \orgname{CIEMAT}, \orgaddress{\city{Madrid}, \postcode{28040}, \country{Spain}}}

\affil[3]{\orgdiv{Department of Mechanical and Aerospace Engineering}, \orgname{Princeton University}, \orgaddress{\city{Princeton}, \postcode{08540}, \state{NJ}, \country{USA}}}

\affil[4]{\orgdiv{Shenzhen Key Laboratory of Ultraintense Laser and Advanced Material Technology, Center for Advanced Material Diagnostic Technology, and College of Engineering Physics}, \orgname{Shenzhen Technology University}, \orgaddress{\city{Shenzhen}, \postcode{518118}, \state{Guangdong}, \country{China}}}


\abstract{
Stellarators confine fusion plasmas using three-dimensional magnetic fields composed of nested toroidal magnetic surfaces. In generic stellarators, trapped particles can drift across these surfaces and degrade plasma confinement. Certain topological properties of the magnetic field strength can suppress these drifts. However, conventional stellarator design approaches typically enforce restrictive constraints to realize such properties, thereby segmenting and limiting the accessible configuration space. In this work, we reformulate the conditions for efficient confinement as constraints on a homeomorphic straightening transformation of the field contours. Within this framework, the various families of stellarator magnetic fields optimized for plasma confinement arise naturally as specific realizations of a unified mapping. This new perspective provides a significantly more comprehensive description of viable stellarator configurations, enabling systematic exploration of trade-offs among confinement quality, geometric complexity, and engineering requirements. We illustrate this approach by presenting a highly compact stellarator design that nevertheless achieves plasma performance comparable to that of leading reactor-scale designs with much larger aspect ratios.
}

\keywords{nuclear fusion, stellarator, stellarator optimization, omnigenity}



\maketitle

\section{Introduction}\label{sec:intro}

Nuclear fusion offers a clean and abundant energy source.
The stellarator \cite{Spitzer1958} was among the earliest concepts proposed for magnetic confinement fusion. 
Unlike tokamaks, stellarators rely on externally generated three-dimensional magnetic fields, enabling intrinsically steady-state operation while avoiding the current-driven disruptions \cite{Kates-Harbeck2019}.
A central challenge for stellarators arises from the lack of continuous symmetry in their magnetic geometry. 
In generic configurations, trapped particles undergo radial drifts across magnetic surfaces, leading to large neoclassical transport and degraded confinement.
Wendelstein 7-X (W7-X) \cite{Grieger1992}, the largest superconducting stellarator in operation, has demonstrated reduced neoclassical transport and achieved record values of the fusion triple product \cite{beidlerDemonstrationReducedNeoclassical2021}. 
These results have renewed interest in stellarators as a viable route to fusion energy.

Neoclassical transport can be reduced if the magnetic-field strength $B=|\mathbf{B}|$ possesses certain \textit{hidden symmetries}.
The most widely studied realization of such symmetry is \textit{quasisymmetry} (QS) \cite{Boozer1983}. 
In a quasisymmetric magnetic field, the magnetic-field strength on a magnetic surface depends on a single angular coordinate when expressed in Boozer coordinates \cite{Boozer1981}. 
This property implies conservation of the canonical momentum associated with the symmetry direction and leads to excellent confinement properties. 
The relative simplicity of the QS condition has made it particularly attractive for numerical optimization, and many modern stellarator designs have been developed by minimizing symmetry-breaking Fourier modes of $B$ \cite{Nuhrenberg1986, Anderson1995, Zarnstorff2001, Liu2020, landremanMagneticFieldsPrecise2022}. 
However, strict quasisymmetry restricts the magnetic-field spectrum to a single helicity, thereby limiting the accessible design space.

More general symmetry conditions relax this restriction while retaining good particle confinement. 
One important example is \textit{omnigenity} \cite{hallThreedimensionalEquilibriumAnisotropic1975, caryHelicalPlasmaConfinement1997, landremanOmnigenityGeneralizedQuasisymmetrya2012}, in which the orbit-averaged radial drift of all trapped particles vanishes. 
This condition is equivalent to requiring the second adiabatic invariant $\mathcal{J}$ to be independent of the field-line label $\alpha$, \textit{i.e.} $\partial\mathcal{J}/\partial\alpha=0$ \cite{helanderTheoryPlasmaConfinement2014}. 
Additional generalisations include \textit{pseudosymmetry} \cite{Isaev1999} and \textit{piecewise omnigenity} \cite{Velasco2024}, which relax different aspects of the omnigenity constraint while preserving favourable confinement properties (as illustrated in Fig.~\ref{fig:intro}). 
Taken together, these concepts suggest that stellarator configurations with good confinement may occupy a much broader region of magnetic-geometry space than strict quasisymmetry alone.

Despite this conceptual generality, optimization of stellarator configurations beyond QS remains considerably more challenging. 
In contrast to quasisymmetric fields, the magnetic-field strength in omnigenous and related configurations does not exhibit straight contours in either equilibrium coordinates or Boozer coordinates. 
As a result, most optimization strategies have relied on indirect objectives \cite{gori1996, Spong2001, Subbotin2006, sanchezQuasiisodynamicConfigurationGood2023, velascoRobustStellaratoroptimization2023} or matching prescribed omnigenous magnetic fields \cite{plunkDirectConstructionOptimized2019, goodmanConstructingPreciselyQuasiisodynamic2023, dudtMagneticFieldsGeneral2024, goodmanQuasiIsodynamicStellaratorsLow2024, helanderOptimisedStellaratorsPositive2024}. 


Here, we introduce a unified geometric framework for stellarator optimization based on a homeomorphic straightening of magnetic-field-strength contours. In this representation, quasisymmetry, omnigenity, pseudosymmetry, and piecewise omnigenity appear as distinct limits of a common symmetry constraint, rather than as separate optimization targets. This formulation replaces pointwise matching of magnetic fields with symmetry-based mode reduction, fundamentally reshaping the optimization landscape.
Using this approach, we obtain stellarator configurations that simultaneously achieve high symmetry precision and geometric compactness. In particular, we demonstrate that omnigenous and piecewise omnigenous configurations with reactor-relevant confinement properties can be realized at aspect ratios as low as four. More broadly, the mapping-based formulation opens previously inaccessible regions of the stellarator design space, enabling new trade-offs among confinement, geometry, and engineering constraints.


\section{Unified optimization framework based on homeomorphic mappings}\label{optimization}

We formulate the optimization of stellarator magnetic fields within a unified framework based on a sequence of coordinate mappings.
Starting from equilibrium code coordinates, the magnetic field is expressed in Boozer coordinates, which provide a natural representation for variations of the magnetic field strength along field lines.
All configurations considered here share this common equilibrium-to-Boozer transformation.

\begin{figure}[!htb]
    \centering
    \includegraphics[width=1\linewidth]{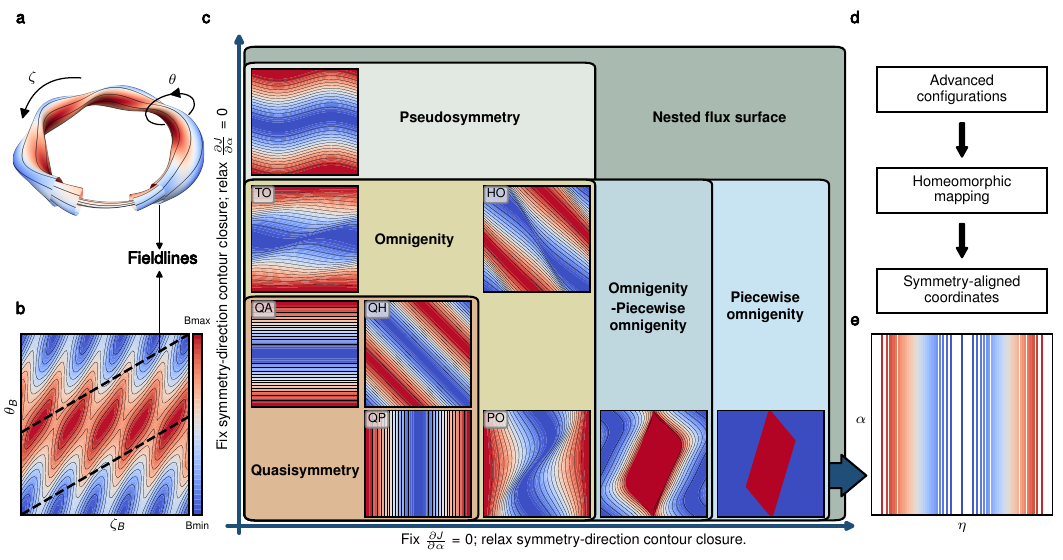}
    \caption{\textbf{Stellarator configurations and the new optimization method.} \textbf{a,} A general stellarator with nested flux surfaces. The colours represent the magnetic field strength $B(s, \theta, \zeta)$, where $s$ is the radial label, $\theta$ is the poloidal angle, and $\zeta$ is the toroidal angle. \textbf{b,} Variation of $B$ in Boozer coordinates $(\theta_B, \zeta_B)$ on a magnetic surface. \textbf{c,} Various stellarator fields that could serve as the basis for reactor candidates. Quasisymmetry is a subset of omnigenity. Omnigenity can be generalized to pseudosymmetry (preserving no locally closed $B$ contours) and piecewise omnigenity (preserving $\partial \mathcal{J} / \partial\alpha = 0$), which can in turn be combined with omnigenity. \textbf{d,} Flow of the new optimization method. \textbf{e,} All configurations can be transformed to symmetry-aligned coordinates with special homeomorphic mappings.}
    \label{fig:intro}
\end{figure}

For magnetic fields with \textit{no} locally closed $B$ contours, including quasisymmetric, omnigenous, and pseudosymmetric configurations, it is possible to introduce symmetry-aligned coordinates $(\alpha,\eta)$ through a homeomorphic mapping in which contours of $B$ become straight along the $\alpha$ direction (Fig.~\ref{fig:intro}).
We denote this mapping by
$
\mathcal{M}: (\alpha,\eta) \rightarrow (\theta_B,\zeta_B).
$
The mapping $\mathcal{M}$ is continuous and invertible, preserving the topological ordering of $B$ contours while allowing geometric deformation.

In this formulation, hidden symmetries are not imposed directly in physical or Boozer coordinates.
Instead, they emerge from constraints on the Boozer-to-$(\alpha,\eta)$ mapping $\mathcal{M}$.
Quasisymmetry corresponds to the limiting case in which the symmetry-aligned coordinates coincide with Boozer coordinates.
Omnigenity, pseudosymmetry, and other hidden symmetries arise when the mapping departs from this limit, while maintaining the absence of locally closed $B$ contours.
From this perspective, these configurations represent geometric regimes within a common mapping space rather than distinct optimization problems.

Given a mapping $\mathcal{M}$, advanced stellarator configurations are optimized by minimizing asymmetric components of $B$ in the symmetry-aligned coordinates.
We define the symmetry objective
\begin{equation}
\label{eq:fsymm}
f_{\mathrm{symm}} = \sum_{m \neq 0} \left( \frac{B_{m,n}}{B_{0,0}} \right)^2 ,
\end{equation}
with
$B(\alpha,\eta) = \sum_{m,n} B_{m,n} \cos(m\alpha - n\eta) ,$
where $B(\alpha,\eta)$ is evaluated by applying the inverse mapping $\mathcal{M}^{-1}$ to the field in Boozer coordinates, together with the appropriate helicity transformation. 
This objective mirrors standard quasisymmetric metrics, but is formulated in a representation that naturally accommodates more general hidden symmetries.

The cost function $f_{\mathrm{symm}}$ can be incorporated into existing stellarator optimization frameworks \cite{stellopt, landremanSIMSOPTFlexibleFramework2021, Dudt2023a}, together with additional objectives and constraints, such as the aspect ratio $A$ and the rotational transform $\iota$.
In the following sections, we show how different mapping constraints give rise to distinct magnetic geometries and confinement properties.


\section{Mapping constraints and emergent physical properties}
Within this framework, different hidden symmetries and confinement properties emerge directly from constraints imposed on $\mathcal{M}$.
Here, we present representative optimized configurations corresponding to quasisymmetric, omnigenous and pseudosymmetric limits.

\subsection{Quasisymmetric mapping as a reference state} \label{sec:qs}
\begin{figure}
    \centering
    \includegraphics[width=1\linewidth]{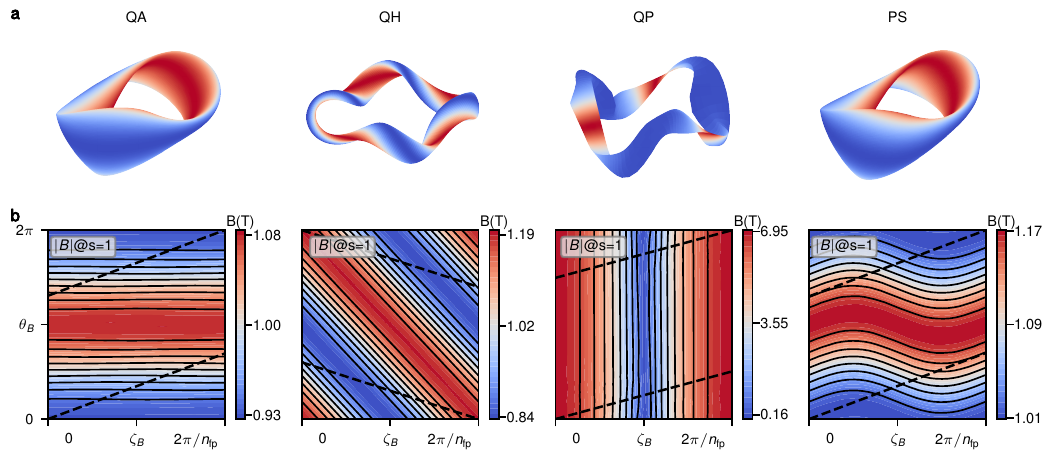}
    \caption{\textbf{Quasisymmetric and pseudosymmetric configurations.} \textbf{a,} 3D plots of the plasma boundary (from left to right: QA, QP, QH, PS). \textbf{b,} $B$ contours in Boozer coordinates on the outermost surface. Dashed lines correspond to the field line.}
    \label{fig:qsps}
\end{figure}

Quasisymmetry corresponds to the case in which symmetry-aligned coordinates coincide with Boozer coordinates. No additional geometric freedom is introduced, and the optimization landscape is comparatively simple.
Three quasisymmetric configurations in different helicities, \textit{quasi-axisymmetry} (QA), \textit{quasi-poloidal} symmetry (QP), and \textit{quasi-helical} symmetry (QH), are shown in Fig.~\ref{fig:qsps}.
These configurations are optimized to be quasisymmetric on the outermost magnetic surface ($s=1.0$, where $s$ is the normalized toroidal flux).
QA has $n_\text{fp}=2$, $A=6$, and nearly constant $\iota=0.70$. 
QP has $n_\text{fp}=3$, $A=6.5$, and $\iota \in [0.76, 1.04]$.
QH has $n_\text{fp}=4$, $A=8$, and nearly constant $\iota=1.20$.
In this work, all configurations are scaled to a volume-averaged field strength of 1 T and a major radius of 1 m.

\begin{figure}
    \centering
    \begin{subfigure}[t]{0.49\linewidth}
        \centering
        \begin{overpic}[width=\linewidth]{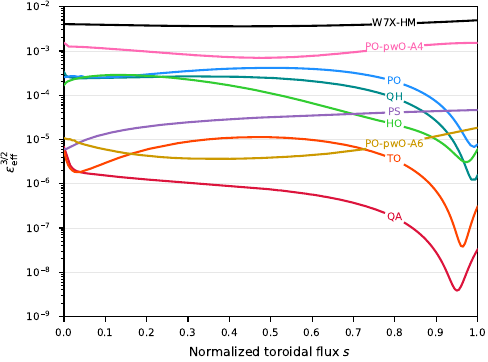}
            \put(3,75){\fontsize{8}{8}\selectfont\sffamily\bfseries a}
        \end{overpic}
    \end{subfigure}\hfill
    \begin{subfigure}[t]{0.49\linewidth}
        \centering
        \begin{overpic}[width=\linewidth]{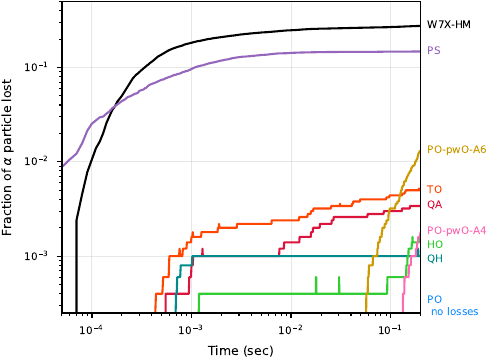}
            \put(3,75){\fontsize{8}{8}\selectfont\sffamily\bfseries b}
        \end{overpic}
    \end{subfigure}
    \caption{\textbf{Confinement properties of all configurations.} \textbf{a,} Neoclassical transport coefficient. \textbf{b,} Collisionless loss fractions of alpha particles at reactor scale.}
    \label{fig:confinement}
\end{figure}

In Boozer coordinates, the guiding-center Lagrangian depends only on $B$, so the canonical momentum is conserved in quasisymmetric fields, yielding good particle confinement.
As shown in Fig.~\ref{fig:confinement}~(a), neoclassical transport in the $1/\nu$ regime, measured by the effective ripple $\epsilon_{\text{eff}}^{3/2}$ calculated using the \texttt{NEO} code \cite{Nemov1999}, is low.
The confinement of fusion-produced alpha particles is evaluated using the \texttt{SIMPLE} code \cite{albert_symplectic_2020} at the ARIES-CS reactor scale \cite{Ku2008} (minor radius of 1.7 m, average field strength of 5.7 T), and 5000 alpha particles (3.5 MeV, isotropic, initialised at $s=0.25$) are followed for 0.2 seconds without collisions.
As shown in Fig.~\ref{fig:confinement}~(b), loss fractions are 0.10\% (QA) and 0.34\% (QH).


Unlike QA and QH, exact QP is impossible near the axis \cite{helanderTheoryPlasmaConfinement2014}.
Although, in principle, it is possible to find a precise QP on a single surface, optimization remains a challenge.
Low-QP-error configurations reported previously \cite{Madan2026} exhibit either large aspect ratios or strong mirror ratios. 
The QP case considered here shows similar limitations and is therefore omitted from the confinement comparison.

\subsection{Omnigenous mappings for extended geometric freedom} \label{sec:omni}
\begin{figure}
    \centering
    \includegraphics[width=\linewidth]{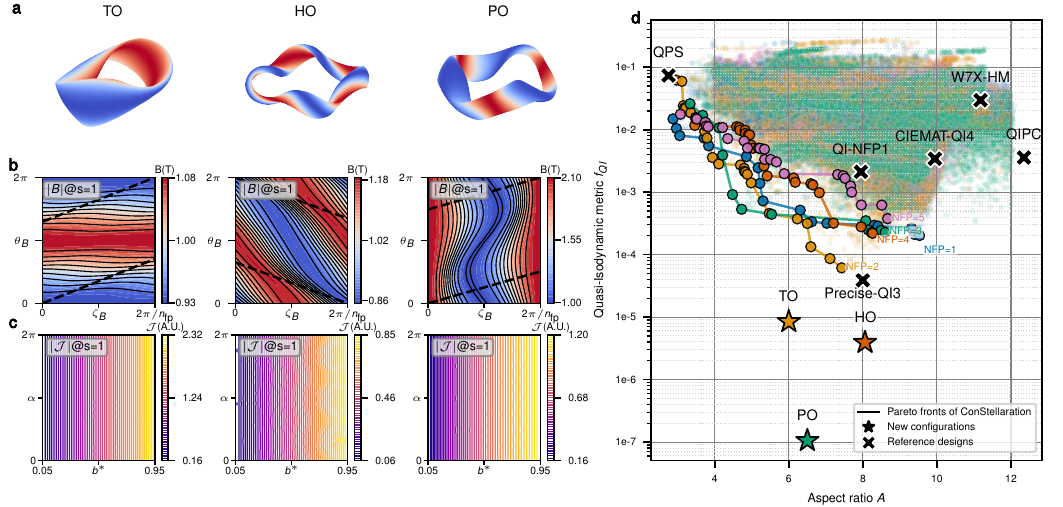}
    \caption{\textbf{Precisely omnigenous configurations.} \textbf{a,} 3D plots of the plasma boundary (from left to right: TO, HO, PO). \textbf{b,} $B$ contours in Boozer coordinates on the outmost surface. \textbf{c,} The distribution of the normalized second adiabatic invariant $\tilde{\mathcal{J}} (\alpha, b^*)$ on the outmost surface ($b^*$ is the relative field strength). \textbf{d,} Comparison of the new omnigenous configurations (stars) with reference QI stellarators (crosses) and the \texttt{ConStellaration} database (dots) in the aspect ratio and QI quality space. The reference QI stellarators include W7X-HM\cite{dinklageMagneticConfigurationEffects2018}, QPS \cite{Spong1998}, QIPC \cite{Subbotin2006}, QI-NFP1\cite{jorgeSinglefieldperiodQuasiisodynamicStellarator2022}, Precise-QI3\cite{goodmanConstructingPreciselyQuasiisodynamic2023}, and CIEMAT-QI4\cite{sanchezQuasiisodynamicConfigurationGood2023}.}
    \label{fig:omni}
\end{figure}

Allowing the mapping $\mathcal{M}$ to deviate from the quasisymmetric limit while enforcing $\partial \mathcal{J}/\partial\alpha = 0$ yields omnigenous configurations. 
In this regime, symmetry is also encoded in the constancy of the second adiabatic invariant on the flux surface.
There are several different omnigenous mappings \cite{caryHelicalPlasmaConfinement1997, landremanOmnigenityGeneralizedQuasisymmetrya2012, dudtMagneticFieldsGeneral2024}.
For omnigenous fields, $B$ contours may close toroidally (TO), poloidally (PO), or helically (HO).
PO, like W7-X, is historically referred to as ``quasi-isodynamicity'' (QI).
Representative omnigenous configurations are shown in Fig.~\ref{fig:omni}.
TO has $n_\text{fp}=2$, $A=6$, and nearly constant $\iota=0.70$. 
PO has $n_\text{fp}=3$, $A=6.5$, and $\iota$ from 0.872 on the magnetic axis to 0.76 at the edge.
HO has $n_\text{fp}=4$, $A=8$, and $\iota \in [1.19, 1.30]$.

In all cases, the normalized second adiabatic invariant $\tilde{\mathcal{J}}$ is almost independent of $\alpha$ (Fig.~\ref{fig:omni}~(c)), confirming precise omnigenity.
Compared with the W7-X high-mirror configuration \cite{dinklageMagneticConfigurationEffects2018}, these designs achieve lower effective ripple and reduced fast-ion losses (Fig.~\ref{fig:confinement}). 
The loss fractions are 0.50\% (TO) and 0.16\% (HO), and are negligible for PO before the slowing-down time.
Importantly, these omnigenous configurations extend beyond previously reported Pareto fronts in the aspect ratio and QI-quality space (Fig.~\ref{fig:omni}(d)).
Reference data-points contain more than 150,000 QI configurations from \texttt{ConStellaration} \cite{Cadena2025} and several QI designs \cite{Spong1998, dinklageMagneticConfigurationEffects2018, dudtMagneticFieldsGeneral2024,sanchezQuasiisodynamicConfigurationGood2023, jorgeSinglefieldperiodQuasiisodynamicStellarator2022, goodmanConstructingPreciselyQuasiisodynamic2023, Subbotin2006}. The mapping-based formulation therefore expands the accessible parameter space of omnigenity. More generally, it indicates that the geometric limits traditionally associated with omnigenous stellarators stem from the way symmetry is formulated and imposed, rather than from the fundamental constraints of three-dimensional magnetic confinement. 

We note that additional stellarator optimization criteria (MHD stability, fast-ion confinement, turbulence, etc) naturally lead to larger $f_{QI}$ and/or $A$. CIEMAT-QI4 \cite{sanchezQuasiisodynamicConfigurationGood2023} is one example of a configuration that is optimized with respect to the standard list of plasma physics criteria required for a stellarator reactor. Moving the Pareto front in the ($f_{QI}$, $A$) space, as done in this work, provides a much-improved starting point for reactor design studies.

\subsection{Pseudosymmetric mappings beyond omnigenity} \label{sec:ps}
If the constraint $\partial \mathcal{J} / \partial\alpha = 0$ is relaxed while preventing locally trapped particles through geometric pitch constraints, pseudosymmetric configurations emerge.
A representative PS configuration ($n_\text{fp}=2$, $A=5$) is shown in Fig.~\ref{fig:qsps}.
The absence of local wells (because $B$ contours close, in this case, toroidally) suppresses neoclassical transport, yielding $\epsilon_\text{eff}^{3/2}$ comparable to TO (Fig.~\ref{fig:confinement}~(a)).
However, because the average radial drift does not vanish, fast-ion confinement is degraded (Fig.~\ref{fig:confinement}~(b)).
This illustrates how distinct mapping constraints produce different confinement properties within the same geometric framework.

\section{Novel configurations combining omnigenity and piecewise omnigenity}\label{pwo}
Omnigenity is often regarded as the minimal symmetry requirement for suppressing bounce-averaged radial drifts of trapped particles. 
However, recent developments have shown that strict global omnigenity is not the only route to good confinement.
Piecewise omnigenity (pwO) relaxes the condition $\partial \mathcal{J} / \partial\alpha = 0$ globally, requiring instead that the second adiabatic invariant satisfies this condition within distinct regions of the magnetic surface \cite{Velasco2024, Velasco2025, Calvo2025}.
In such configurations, locally-closed $B$ contours are permitted. In the most extreme case of these families of optimized fields, $B$ takes to values $B_\text{max}$ within a parallelogram-shaped region, and $B_\text{min}$  outside it. While such extreme examples are compatible with other reactor design constraints~\cite{Fernandez-Pacheco2026}, they display a large aspect ratio. Other possibilities, in which omnigenity is combined with piecewise omnigenity, have been proposed \cite{Velasco2025}, but are inaccessible with the design approach of \cite{Fernandez-Pacheco2026}.

Within the mapping-based framework introduced here, pwO features emerge naturally when global omnigenity constraints are partially relaxed. 
In the construction of omnigenous mappings, symmetry-aligned coordinates are typically defined over the full angular domain $\eta \in [-\pi, \pi)$.
If the mapping is shaped such that $B$ contours are strongly deformed and the range of $\eta$ is limited, the $B_{\text{max}}$ contours may close locally with the approximate shape of a parallelogram.
At the same time, $B_{\text{min}}$ regions retain omnigenous and smooth variation of $B$ ensures continuity of good particle confinement.
In these locally closed regions, multiple values of the second invariant coexist while approximately satisfying the pwO condition~\cite{Velasco2024,Velasco2025}.

\begin{figure}
    \centering
    \includegraphics[width=1\linewidth]{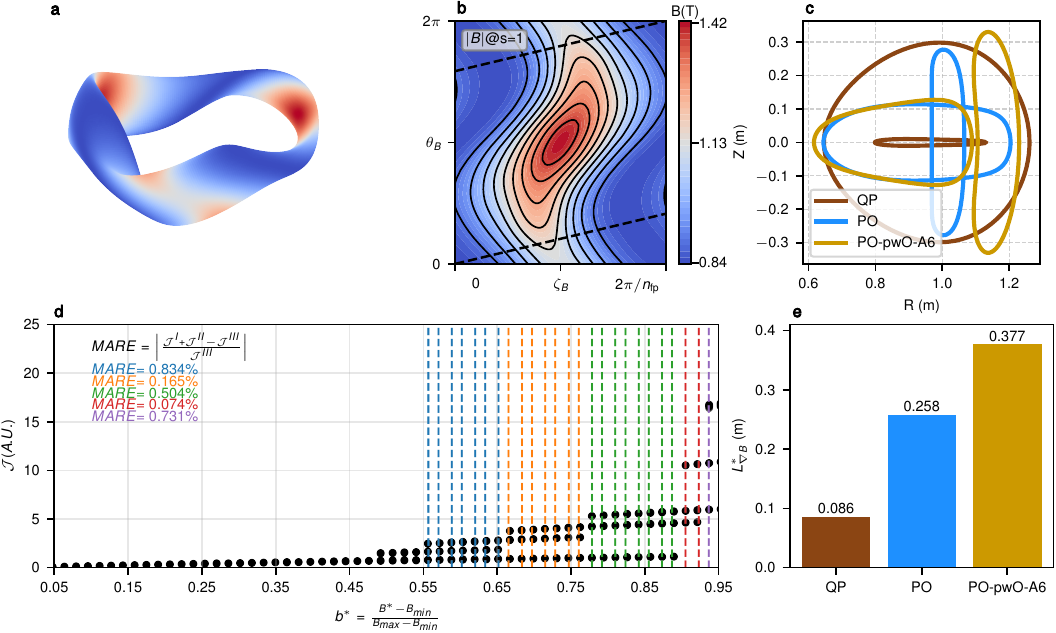}
    \caption{\textbf{The PO-pwO-A6 configuration.} \textbf{a,} 3D plot of the plasma boundary. \textbf{b,} $B$ contours in Boozer coordinates on the outmost surface. \textbf{c,} Cross-sections of QP, PO, and PO-pwO-A6 configurations at toroidal angles of 0 and $\pi/n_\text{fp}$. \textbf{d,} $\mathcal{J}$ on the outmost surface as a function of $b^*$. There are multiple invariant branches in high-field regions. The mean absolute relative error (MARE) is below 1\% in each region. \textbf{e,} The minimum magnetic gradient scale length $L^*_{\nabla B}$ of three configurations.}
    \label{fig:pwo}
\end{figure}

A representative configuration combining omnigenity and piecewise omnigenity is shown in Fig.~\ref{fig:pwo}, denoted as PO-pwO-A6.
It has $n_\text{fp}=3$, $A=6$, and $\iota \in [0.62, 0.71]$.
In this configuration, $B$ contours close locally in the high-field region, while the low-field region behaves as in a PO field (Fig.~\ref{fig:pwo}(b)).
The resulting topology differs from previously constructed pwO equilibria, in which $B_{\text{min}}$ and $B_{\text{max}}$ regions are single-valued \cite{Fernandez-Pacheco2026}. 
Here, the coexistence of multiple discrete values of $\mathcal{J}$ (Fig.~\ref{fig:pwo}(d)) arises directly from the mapping structure rather than from imposed contour closure.
The PO-pwO-A6 equilibrium exhibits low $\epsilon_\text{eff}^{3/2}$ and alpha-particle loss fractions of approximately 1\% at the reactor scale (Fig.~\ref{fig:confinement}).

From a broader perspective, QS can be viewed as a special case within omnigenity. In contrast, omnigenity-piecewise-omnigenity fields represent a partial relaxation of global omnigenity within the same unified framework.
Relaxing constraints may improve engineering feasibility.
In Fig.~\ref{fig:pwo}(c), this is illustrated by showing the cross-sections of QP, PO, and PO-pwO-A6.
The three configurations have the same $n_\text{fp}$ with comparable $\iota$ and $A$.
However, PO-pwO-A6 is substantially less elongated than QP.
The maximum elongation, which is associated with coil difficulty \cite{hudsonDifferentiatingShapeStellarator2018}, is 24 for QP and 5.0 for PO-pwO-A6.
The minimum magnetic gradient scale length $L^*_{\nabla B}$ \cite{Kappel_2024}, another commonly used coil complexity proxy \cite{Lion2025}, is also improved for PO-pwO-A6 (Fig.~\ref{fig:pwo}(e)).
These geometric features arise from the additional degrees of freedom afforded by formulating symmetry in the mapping space rather than enforcing global contour closure.

\section{Ultimately compact reactor design}
Stellarator reactors are expected to have good performance in several aspects \cite{Gates2018}, such as reduced neoclassical transport, MHD stability at finite beta, good alpha-particle confinement, and low turbulence transport.
QI (PO) configurations are currently among the leading candidates for stellarator reactors \cite{sanchezQuasiisodynamicConfigurationGood2023, Warmer2024, Lion2025, Hegna2025}, yet most reactor-scale realizations exhibit large aspect ratios ($>10$), reflecting a perceived trade-off between symmetry precision and geometric compactness.
Within the mapping-based formulation, this trade-off is substantially relaxed. 
Because symmetry constraints are imposed through mode reduction in symmetry-aligned coordinates rather than through global contour closure, the accessible configuration space expands. 
Compact equilibria with high omnigenity precision become attainable without sacrificing confinement quality (Fig~\ref{fig:omni}(d)).
By relaxing the constraint on global omnigenity, PO-pwO configurations provide extra freedom in geometries while preserving good confinement.

\begin{figure}
    \centering
    \includegraphics[width=1\linewidth]{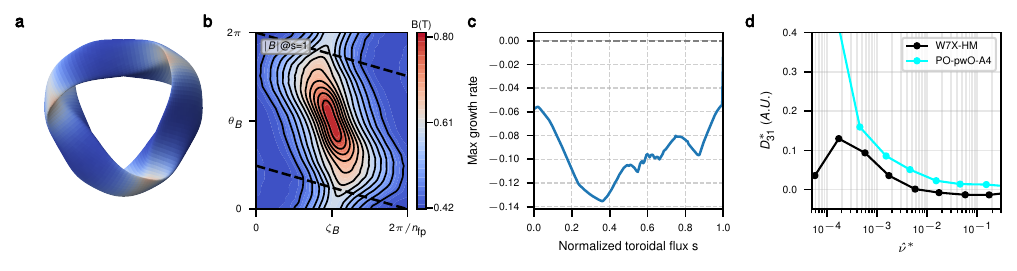}
    \caption{\textbf{Compact PO-pwO reactor design.} \textbf{a,} 3D plot of the plasma boundary. \textbf{b,} $B$ contours in Boozer coordinates on the outmost surface. \textbf{c,} Infinite-n ballooning mode growth rate across radial flux surfaces. \textbf{d,} The bootstrap transport coefficient $D^*_{31}$ of the PO-pwO-A4 configuration and W7X-HM.}
    \label{fig:reactor}
\end{figure}

Fig~\ref{fig:reactor} presents a three-period PO-pwO configuration with aspect ratio $A=4$.
Despite its compactness, the configuration retains effective ripple $\epsilon_\text{eff}^{3/2}$ of order $10^{-4}$ and alpha-particle loss fraction below one percent (0.18\%) at reactor scale (Fig.~\ref{fig:confinement}).
The equilibrium possesses a vacuum magnetic well and remains ballooning-stable at $\beta=3.0\%$.
The bootstrap transport coefficient, $D^*_{31}$, is comparable to that of W7-X in the reactor-relevant collisionality regime, implying small bootstrap currents.
The edge rotational transform $\iota \approx 0.75$ is compatible with island divertors.

This configuration illustrates that high symmetry precision and compact aspect ratio are not intrinsically incompatible. 
It results from the formulation of piecewise omnigenity within the framework of the method developed in this work.
By reformulating hidden symmetries in mapping space, the accessible trade-offs between confinement and geometry are fundamentally reshaped.

\section{Discussion and outlook}\label{outlook}
Hidden symmetry in stellarators is traditionally formulated through conditions imposed directly on the magnetic field or on particle invariants.
The mapping-based formulation introduced here shifts the focus from the magnetic field itself to the coordinate transformation through which symmetry is expressed. 
In this setting, symmetry conditions correspond to restrictions on a homeomorphic mapping between Boozer coordinates and symmetry-aligned coordinates.
By enforcing symmetry through mode reduction in symmetry-aligned coordinates, the admissible design space changes, enabling compact equilibria with high symmetry precision.
The new approach opens avenues for deeper exploration within the design space for future stellarator reactors. 

It is worth noting that coordinate-based perspective opens the possibility of designing stellarators that, resembling W7-X, overcome its limitations. 
Although it is nominally considered QI, W7-X was obtained through direct optimization of radial drift rather than through explicit enforcement of quasi-isodynamicity. 
For this reason, the structure of its $B$ contours shares qualitative features with PO-pwO~\cite{Velasco2025}. This observation is consistent with the fact that W7-X achieves reactor-relevant neoclassical confinement despite deviations from strict quasi-isodynamicity. Crucially, it displays only moderate elongation.
The PO-pwO concept will be further explored in other work \cite{Velasco2026, Liu2026}

Because symmetry is controlled through the mapping, the same formalism can be extended naturally to more realistic design requirements. 
The symmetry objective can be applied to multiple flux surfaces, achieving global symmetries.
It can be combined with other metrics to accommodate additional requirements, such as turbulence transport \cite{mynickOptimizingStellaratorsTurbulent2010, xanthopoulosControllingTurbulencePresent2014, Proll2015, goodmanQuasiIsodynamicStellaratorsLow2024} and coil realizability \cite{focus}.
The mapping space itself is unlikely to be exhausted by the constraint families considered here. 
For example, generalized omnigenous constructions that relax the equality of local extrema while preserving bounce-averaged drift suppression \cite{parraLessConstrainedOmnigeneous2015} suggest additional omnigenity mappings.
Additional configuration families remain to be systematically explored within the expanded mapping space. 
In particular, toroidal and helical omnigenous regimes, viewed as generalized extensions of quasi-axisymmetric and quasi-helical symmetry, may provide further geometric flexibility.

\section*{Methods}




\textbf{VMEC.}
MHD equilibria were computed using the Variational Moments Equilibrium Code (VMEC) \cite{hirshmanSteepestdescentMomentMethod1983}, which solves the ideal-MHD force-balance system under the assumption of nested flux surfaces via minimization of the total plasma energy. The plasma boundary and flux-surface geometry are represented spectrally in poloidal and toroidal angles. Fixed-boundary equilibria were obtained by prescribing boundary Fourier coefficients.

\textbf{Aspect ratio.}
The aspect ratio is defined as $A=R_{\mathrm{major}}/a$, where for the LCFS we set
$a=\sqrt{\langle A_{\mathrm{cs}}\rangle/\pi}$ from the toroidally averaged poloidal cross-sectional area and
$R_{\mathrm{major}}=V/(2\pi^2 a^2)$ from the enclosed volume $V$.

\textbf{Elongation.}
Flux-surface elongation was evaluated from boundary cross-sections obtained by intersecting the LCFS with planes normal to the magnetic axis at uniformly sampled toroidal angles. For each seed angle $\phi_0$, the cutting plane passes through the magnetic-axis point $\mathbf{p}_{\mathrm{ax}}(\phi_0)$ and has normal given by the unit tangent of the magnetic axis $\hat{\mathbf{t}}(\phi_0)$; the intersection curve is constructed by sampling poloidal angles $\theta$ and solving for $\phi(\theta)$ such that $\hat{\mathbf{t}}(\phi_0)\!\cdot\!\big(\mathbf{r}(\theta,\phi(\theta))-\mathbf{p}_{\mathrm{ax}}(\phi_0)\big)=0$. The resulting planar curves were approximated as polygons, from which area $A$ and perimeter $P$ were computed. Because the cross-sections are not generally elliptical, we define an equivalent ellipse with identical area and perimeter,
$A=\pi ab$ and $P=4a\,E(m)$, where $m=1-(b/a)^2$ and $E$ is the complete elliptic integral of the second kind. The semi-axes $(a,b)$ are obtained numerically, and elongation is defined as $\kappa=\max(a,b)/\min(a,b)$. We report the maximum elongation over all sampled cross-sections.

\textbf{VMEC–SIMSOPT optimization.}
Optimizations were performed using SIMSOPT \cite{landremanSIMSOPTFlexibleFramework2021} with VMEC as the MHD equilibrium solver. 
Boundary Fourier coefficients were used as design variables. 
A continuation strategy was used in which VMEC resolution and boundary spectral content were progressively increased.
Unless otherwise stated, runs were initialized at $(\mathrm{MPOL},\mathrm{NTOR},\mathrm{NS})=(3,3,50)$ and the boundary truncation followed $m_{\max}=\mathrm{MPOL}-2$ and $n_{\max}=\mathrm{NTOR}-2$. 
For the TO/HO families, baseline QA/QH equilibria were first obtained at $\mathrm{FTOL}=10^{-11}$ and then continued through $(\mathrm{MPOL},\mathrm{NTOR})=(5,5)$ (boundary $m_{\max}=n_{\max}=3$) and $(12,12)$ (boundary modes increased to $7$), in separate runs used to refine QA/QH and to initialize TO/HO. For PO, PS, and PO-pwO-A6, we used analogous continuation schedules: PO $(\mathrm{FTOL}=10^{-14},\,\mathrm{MPOL}/\mathrm{NTOR}:3/3\rightarrow13/13,\,n_{\max}/m_{\max}:1\rightarrow7)$; PS $(\mathrm{FTOL}=10^{-14},\,3/3\rightarrow9/9,\,n_{\max}/m_{\max}:1\rightarrow5)$; PO-pwO-A6 $(\mathrm{FTOL}=10^{-15},\,4/4\rightarrow6/6,\,n_{\max}/m_{\max}:2\rightarrow4)$, warm-started from a PO equilibrium truncated to $m_{\max}=n_{\max}=2$ and optimized on $s=\{0.4,0.7,1.0\}$ with internal-surface mapping degrees of freedom enabled and an elongation constraint $\kappa\le5$.

\textbf{DESC design iteration.}
The PO–pwO–A4 configuration was developed using DESC \cite{DESC_Part0,DESC_PartI,DESC_PartII,DESC_PartIII}, which enables GPU-accelerated automatic differentiation. The optimization included explicit constraint terms targeting ballooning stability, magnetic well depth, and shaping metrics, like elongation.

\textbf{Homeomorphic Mappings.}
In the omnigenity mapping, we introduce an intermediate coordinate system
$(\tilde{\theta},\tilde{\zeta})$ to construct a transformation from
$(\alpha,\eta)$ to Boozer angles $(\theta_B,\zeta_B)$.
We set $\tilde{\theta}=\alpha$ and define
\begin{equation}
\label{eq:mapping}
\tilde{\zeta}(\alpha,\eta)=\eta - D(\eta)\,S(\alpha,\eta),
\end{equation}
where $\alpha\in[0,2\pi)$ labels field lines and $\eta\in[-\pi,\pi)$ labels
$B$-contours ($\eta=0$ at minimum $B$ and $\eta=\pm\pi$ at maximum $B$).
Here, $S(\alpha,\eta)$ controls contour shape, while $D(\eta)$ specifies the
bounce distance between equal-$B$ contours along field lines.
To preserve stellarator symmetry, we take $S(\alpha,\eta)$ to be odd in its
argument and $D(\eta)$ to be even in $\eta$, and represent them using Fourier
series
\begin{equation}
\label{eq:SD}
\left\{
\begin{aligned}
S(\alpha,\eta) &= \sum_m s_m \sin\!\left[m\,y(\alpha,\eta)\right],\\
D(\eta) &= \pi-|\eta|+\sum_n d_n \cos\!\left[\left(n+\tfrac12\right)\eta\right],
\end{aligned}
\right.
\end{equation}
with Fourier coefficients $\{s_m\}$ and $\{d_n\}$.
The phase variable
\begin{equation}
y(\alpha,\eta)=\mathbf{Y}\,[\alpha\ \eta]^{\mathrm T}
\end{equation}
selects the helicity of the mapping. In this work, we use
\begin{equation}
\mathbf{Y}_{\mathrm{TO}}=
\begin{bmatrix} n_{\mathrm{fp}} & -{n_{\mathrm{fp}}}/{\iota}\end{bmatrix},\quad
\mathbf{Y}_{\mathrm{PO}}=
\begin{bmatrix} 1 & -{\iota}/{n_{\mathrm{fp}}}\end{bmatrix},\quad
\mathbf{Y}_{\mathrm{HO}}=
\begin{bmatrix} 1 & {\iota}/{(\iota+n_{\mathrm{fp}})}\end{bmatrix},
\end{equation}
for toroidal omnigenity (TO), poloidal omnigenity (PO), and helical omnigenity
(HO), respectively, where $\iota$ is the rotational transform and
$n_{\mathrm{fp}}$ is the field periodicity.
The Boozer angles with field periodicity are obtained via a linear map
\begin{equation}
\label{eq:thetazeta}
[\theta_B\ \zeta_B]^{\mathrm T}=\mathbf{M}\,[\tilde{\theta}\ \tilde{\zeta}]^{\mathrm T},
\end{equation}
with
\begin{equation}
\label{eq:T}
\mathbf{M}_{\mathrm{TO}}=
\begin{bmatrix}0 & 1 \\ 1 & 0\end{bmatrix},\quad
\mathbf{M}_{\mathrm{PO}}=
\begin{bmatrix}1 & 0 \\ 0 & 1/n_{\mathrm{fp}}\end{bmatrix},\quad
\mathbf{M}_{\mathrm{HO}}=
\begin{bmatrix}1 & 0 \\ -1/n_{\mathrm{fp}} & -1/n_{\mathrm{fp}}\end{bmatrix},
\end{equation}
for TO, PO, and HO, respectively.
Note that the $1/n_{\mathrm{fp}}$ scaling is applied in PO and HO (but not TO)
to avoid rotational transforms comparable to $n_{\mathrm{fp}}$ in large-$n_{\mathrm{fp}}$
configurations.

For pseudosymmetry (PS), the bounce distance is not constrained, so we use a
more flexible monotonic mapping.
For piecewise omnigenity (pwO), we modify Eq.~\eqref{eq:mapping} to
\begin{equation}
\tilde{\zeta}
= \eta^\ast - (\pi-|\eta|)\,s_1\sin\!\left[y(\alpha,\eta^\ast)\right],
\end{equation}
where
\begin{equation}
\eta^\ast=\eta\left(p|\eta|-p\pi+1\right),
\end{equation}
and $p$ is a numerical factor controlling the remapping of $\eta$.

For the configurations shown in this paper, we used the minimal
single-mode parameterization with $d_n=0$ (so $D(\eta)=\pi-|\eta|$) and only one
non-zero coefficient in $S$.
Specifically, $S=0$ for QA/QP/ QH, $S=0.03\sin y$ for TO, $S=0.3\sin y$ for PO, and
$S=-0.15\sin y$ for HO.
For PS we used $\theta_B=\eta-0.08\pi\sin\alpha$ and $\zeta_B=\alpha/n_{\mathrm{fp}}$.
For PO-pwO-A6 we used $s_1=0.4$, $p=-0.6$, and restricted $\eta\in[-0.225\pi,\,0.225\pi]$.

\textbf{Final high-resolution verification.}
Final equilibria were verified with higher radial and spectral resolutions to ensure convergence. The verification runs used: TO $(\mathrm{MPOL},\mathrm{NTOR},\mathrm{NS},\mathrm{FTOL})=(16,16,201,10^{-16})$; HO: $(13,16,201,10^{-16})$; PO: $(13,13,121,10^{-16})$; PS: $(14,14,201,10^{-14})$; PO-pwO-A6: $(12,12,201,2\times10^{-17})$; and PO-pwO-A4: $(11,10,176,10^{-16})$.

\textbf{Quasi-isodynamic metric $f_{QI}$.}
Departure from quasi-isodynamicity was quantified using the QI metric of Goodman \textit{et al.} \cite{goodmanConstructingPreciselyQuasiisodynamic2023}, consistent with the \texttt{ConStellaration} database \cite{Cadena2025}. On the LCFS ($s=1$), the metric is the normalized mean-square deviation between the magnetic field strength $\tilde B(\alpha,\varphi)$ in Boozer coordinates and its perfectly-QI reference $\tilde B_{\mathrm{QI}}$ constructed from the sampled wells via the squash--stretch--shuffle procedure,
\begin{equation}
f_{QI}(s)=\frac{n_{\mathrm{fp}}}{4\pi^2}\,\frac{1}{\big(B_{\max}-B_{\min}\big)^2}
\int_{0}^{2\pi}\!\! d\alpha \int_{0}^{2\pi/n_{\mathrm{fp}}}\!\! d\varphi\,
\Big(\tilde B(s,\alpha,\varphi)-\tilde B_{\mathrm{QI}}(s,\alpha,\varphi)\Big)^2 .
\end{equation}
Specifically, we sample $n_\alpha={75}$ field lines uniformly in $\alpha\in[0,2\pi)$ and $n_\varphi={601}$ points with $n_{Bj}={401}$ magnetic field levels between the bottom and top of the magnetic well in $\varphi\in[0,2\pi/n_{\mathrm{fp}}]$ on the outermost surface ($s={1.0}$).

\textbf{Magnetic well.}
The magnetic well is characterized using the specific volume
$U(\psi_t)\equiv dV/d\psi_t=\int dl/|B|$, for which $dU/d\psi_t<0$ indicates a magnetic well (and $dU/d\psi_t>0$ a magnetic hill). The well-depth is quantified as $-\Delta U/U \equiv (U_{\mathrm{axis}}-U(\rho))/U_{\mathrm{axis}}$, so that increasing $-\Delta U/U$ with radius corresponds to a well region.


\textbf{Ballooning stability.}
Ideal MHD ballooning stability was evaluated using COBRAVMEC, which applies the infinite-(n) ballooning formalism to VMEC equilibria. Negative eigenvalues indicate instability.
COBRAVMEC was run using \texttt{NTHETA}={12} and \texttt{NZETA}={12} starting angles in $(\theta_0,\zeta_0)$, with \texttt{$k_w$}={10}. 
Ballooning growth rates were evaluated on all flux surfaces.

\textbf{MONKES.}
Monoenergetic neoclassical transport coefficients were computed using MONKES\cite{Escoto_2024} on the outermost flux surface. The normalized coefficients ($D^*_{11}$ and $D^*_{31}$) were obtained from the monoenergetic coefficients via standard normalization factors involving the device geometry and rotational transform.
\texttt{MONKES} was run on the flux surface $s=1.0$ with a prescribed radial-electric-field grid \texttt{$E_r$=0}, grid resolutions were chosen \texttt{$N_{theta}$} = 120,\texttt{$N_{zeta}$} = 120 and \texttt{$N_{\xi}$} = 120.


\textbf{Minimal magnetic gradient scale length $L_{\nabla B}^*$.}
The magnetic gradient scale length \cite{Kappel_2024} was defined as
\begin{equation}
L_{\nabla B}=\frac{\sqrt{2}\,B}{\left\|\nabla\mathbf{B}\right\|_{F}},\qquad 
\left\|\nabla\mathbf{B}\right\|_{F}=\left(\sum_{i,j}\left(\frac{\partial B_i}{\partial x_j}\right)^2\right)^{1/2},
\end{equation}
which serves as a proxy for the magnetic-field variation scale relevant to coil realizability.
The minimal magnetic gradient scale length $L_{\nabla B}^*$ is the smallest value of $L_{\nabla B}$ on the outermost surface.


\backmatter

\bmhead{Acknowledgements}

This work was supported by the Strategic Priority Research Program of the Chinese Academy of Sciences under Grant No. XDB0790302, by the National Natural Science Foundation of China (NSFC) with Grant No. 12475229, and the Anhui Provincial Key Research and Development Project under Grant No. 2023a05020008. This research was partially supported by grants PID2021-123175NB-I00 and PID2024-155558OB-I00, Ministerio de Ciencia, Innovaci\'on y Universidades, Spain.

\section*{Declarations}


\begin{itemize}

\item \textbf{Competing interests:} 
The authors declare no competing interests.

\item \textbf{Data availability :} 
The data supporting the findings of this study will be available at Zenodo: Liu, H. (2026). Data for paper "Optimizing stellarators with hidden symmetry" [Data set]. Zenodo. https://doi.org/10.5281/zenodo.19068436

\item \textbf{Code availability:} 
The code implementation described in this work is available at \texttt{https://github.com/USTCstellarators/OOPS.git}.

\item \textbf{Author contribution:}
H.L., G.Y., and C.Z. conceived and designed the study.
H.L. and G.Y. led the implementation and integration of the code.
H.L. and C.Z. wrote the initial draft of the manuscript.
C.Z., J.L.V., and M.Y. substantially revised the manuscript.
R.G., D.P., and E.K. contributed to code integration, migration, and technical troubleshooting.
W.D., S.W., and G.Z. reviewed the manuscript and contributed to the discussion and interpretation of the results.
All authors discussed the results and approved the final manuscript.
\end{itemize}



\bibliography{ref}

\end{document}